# Optical Scintillation Imaging for Dynamic Source Tracking Dosimetry During Prostate HDR Brachytherapy

**Kevin Willy[1*], Megan Clark[4], Petr Bruza[1], Shauna McVorran[5], and David Gladstone[1,2,3]**

1 Thayer School of Engineering at Dartmouth College, Hanover, NH 03755

2 Dartmouth Cancer Center, Lebanon, NH 03756

3 Department of Radiation Oncology and Applied Sciences, Geisel School of Medicine, Dartmouth Lebanon, NH 03756

4 Stanford Medicine Department of Radiation Oncology, Stanford, CA 94305

5 University of Washington Medicine's Fred Hutch Cancer Center, Seattle, WA 98195

*Author to whom any correspondence should be addressed.

E-mail: kevin.j.willy.th@dartmouth.edu

## Author Contribution Statement

**Kevin Willy**: Conceptualization, methodology, software, validation, formal analysis, investigation, data curation, writing-original draft, review & editing, visualization. **Megan Clark**: conceptualization, writing- review & editing**. Petr Bruza**: Supervision, writing – editing & review, **Shauna McVorran**: conceptualization, funding acquisition, writing – review & editing. **David Gladstone**: supervision, funding acquisition, writing, review, and editing.

## Acknowledgements

This work was funded by the American Cancer Society (ACS): IRG-21-136-36-IRG-07 and the National Institutes of Health (NIH) through the National Cancer Institute (NCI) Cancer Center Support Grant: P30CA023108. This device is patent-pending, application No. 63444926: Transrectal Probe for Scintillation Imaging as a Method of Real Time Dosimetry of High-Dose-Rate Brachytherapy.

### Conflict of Interest Statement

The authors have no relevant conflicts of interest to disclose.

## Abstract

**Background:** Brachytherapy enables highly conformal radiotherapy, but its efficacy relies on surgical precision. Despite risks of catheter displacement or source-indexing errors, current clinical workflows lack real-time in vivo dosimetry (IVD) systems that verify dose delivery without disrupting the procedure.

**Purpose:** This study characterizes a novel, scintillator-based optical dosimetry system integrated into a transrectal probe providing tracking of source position, speed, dose-rate, and time-resolved cumulative dose for high-dose-rate (HDR) brachytherapy.

**Methods:** The dosimeter utilizes a 67 mm x 5 mm line scintillator imaged by a CMOS camera through a 30-degree endoscopic lens. Geometric transformations and flat field corrections were implemented to account for the barrel distortion and lens vignetting, respectively. Absolute dose calibration was established using a TOPAS Monte Carlo model that incorporated the probe's specific air-cavity geometry. The system was validated in a water-filled phantom using an Iridium-192 (Ir192) source, with needles placed at radial offsets of 7, 10, and 15 mm.

**Results:** The dosimeter demonstrated high spatial accuracy with variation of ±2 % across the 67 mm length. High linearity was observed across the clinical dose range of 5 cGy/sec – 21 cGy/sec tested ($\boldsymbol{R^2} \geq \mathbf{0.995}$). The Monte Carlo model revealed a 3.75% dose-rate reduction compared to the standard TG-43 water-equivalent conditions. In phantom testing, the system successfully resolved a 3 mm physical needle displacement, which was subsequently verified by computed tomography (CT). Measured dose-rate profiles showed high concordance with Monte Carlo peak magnitude and spatial distribution, and total integral dose measurements agreed within 1%.

**Conclusions:** This endoscopic optical dosimetry system provides high-resolution, time-resolved verification of HDR brachytherapy parameters. By mapping dose data directly to the measurement plane within a standard ultrasound transducer, this technology offers a clinical tool for detecting delivery errors and improving patient safety with minimal modifications to existing treatment protocols.

## 1. Introduction

Brachytherapy enables highly conformal radiotherapy delivery by placing a radioactive source directly within or against the treatment target tissue. Brachytherapy treats target tissue while sparing adjacent organs at risk due to the rapid dose fall-off. However, this high gradient dose delivery, requires accurate HDR source positioning, since minor deviations may result in underdosing of the primary target or overdosing of the surrounding healthy tissue.[1] Currently, this challenge is addressed by using remote afterloader systems in which the radioactive seed is attached to a guide wire, and a stepper motor translates the wire along a guide path toward the predefined dwell positions within the chosen applicator. Dwell positions and times may be chosen to account for the achieved applicator placement in the patient, thereby arriving at an optimal dose distribution. Although afterloader systems have been shown to maintain high placement accuracy, the space within the guide catheter that ensures the wire has sufficient clearance for safe traversal also results in local wire sag and may introduce <1mm deviations in dwell spot positioning, thereby altering the delivered dose distribution. Furthermore, the stepper motor operating cycle requires acceleration and deceleration stages between dwell positions, which results in dose profiles that cannot be accurately accounted for in the treatment planning software, leading to further deviation in the delivered dose.[1–3] In-vivo dosimetry is desired to improve treatment quality assurance by monitoring and recording delivered treatments, enabling verification of treatment delivery and trend analysis of minor deviations. Past methods of in vivo dosimetry for brachytherapy include multiple-point measurements decomposed to determine source position using various dosimeters[4–8], pinhole optics paired with film or gamma cameras, and flat panel detectors[9,10]. These implementations lacked the spatial and temporal resolution required for real-time source-position monitoring and quantitative post-delivery dosimetric verification.

Optical dosimetry systems have shown success pairing scintillators with camera imaging to measure in vivo dose.[11,12] These optical systems can be flexibly constructed using mirrors or specialty lenses to image in challenging geometric conditions.[13] Overall, the freedom in choice of scintillators, camera sensors, and optics provides opportunities to measure dose in a broad range of treatment modalities. In brachytherapy, camera systems have been used to quantify dose-affecting variables, such as source speed and position. Past methods have employed high-speed cameras to measure source transit speeds[14] and gamma cameras to track source position[15]. However, these systems did not provide simultaneous quantitative dosimetry and real-time positional monitoring within a single clinically integrated platform.

Building on prior optical dosimetry systems, a CMOS camera with an endoscopic lens was developed to image a line scintillator inside a transrectal probe. By capturing the spatiotemporally resolved scintillator output during HDR source irradiation, source position, speed, dwell times, and absolute dose rate were extracted, thereby providing a clinically viable method for in vivo treatment verification without altering standard clinical workflows. The system additionally provides a live video feed, enabling qualitative real-time monitoring during delivery and allowing clinicians to quickly identify gross errors such as a stuck or broken source, incorrect dwell position, or unexpected source motion.

## 2 Methods

### 2.1 System Design

An optical dosimetry probe was designed based on the physical structure of a commercially available transrectal ultrasound probe (Edan Endorectal Transducer E741-2). The inside of the probe features a cavity that houses an endoscopic camera lens with a 30-degree viewing angle, oriented to image a 67 mm by 5 mm rectangular plastic-matrix composite scintillator (Green 600 Penn-Jersey X-ray). The probe and lens are mounted on a CMOS sensor camera module (FLIR Blackfly S) via a focusing adapter (Stryker 30-degree). By orienting the active side of the scintillator to face inward within the probe and placing the strip at the superior part of the guiding surface, the measurement plane was positioned directly beneath the ultrasound probe's transducer, measuring dose near the anterior rectal wall. During clinical delivery, a live video feed provides real-time visualization of scintillator emission.

### 2.2 Test Phantom Design

A water-filled phantom was designed to simulate a rectal cavity and house the probe, with interstitial brachytherapy needles positioned at 7, 10, and 15mm from the needle center to the scintillator's visible surface. Figure 1 shows a computer-aided design (CAD) model cross-section of the transrectal ultrasound, the endoscopic needle mount, scintillator target, and the three sample needle placements within the water phantom. The phantom features a cavity for inserting the probe. All components are suspended in a water bath to simulate the tissue environment during use (Figure 1).

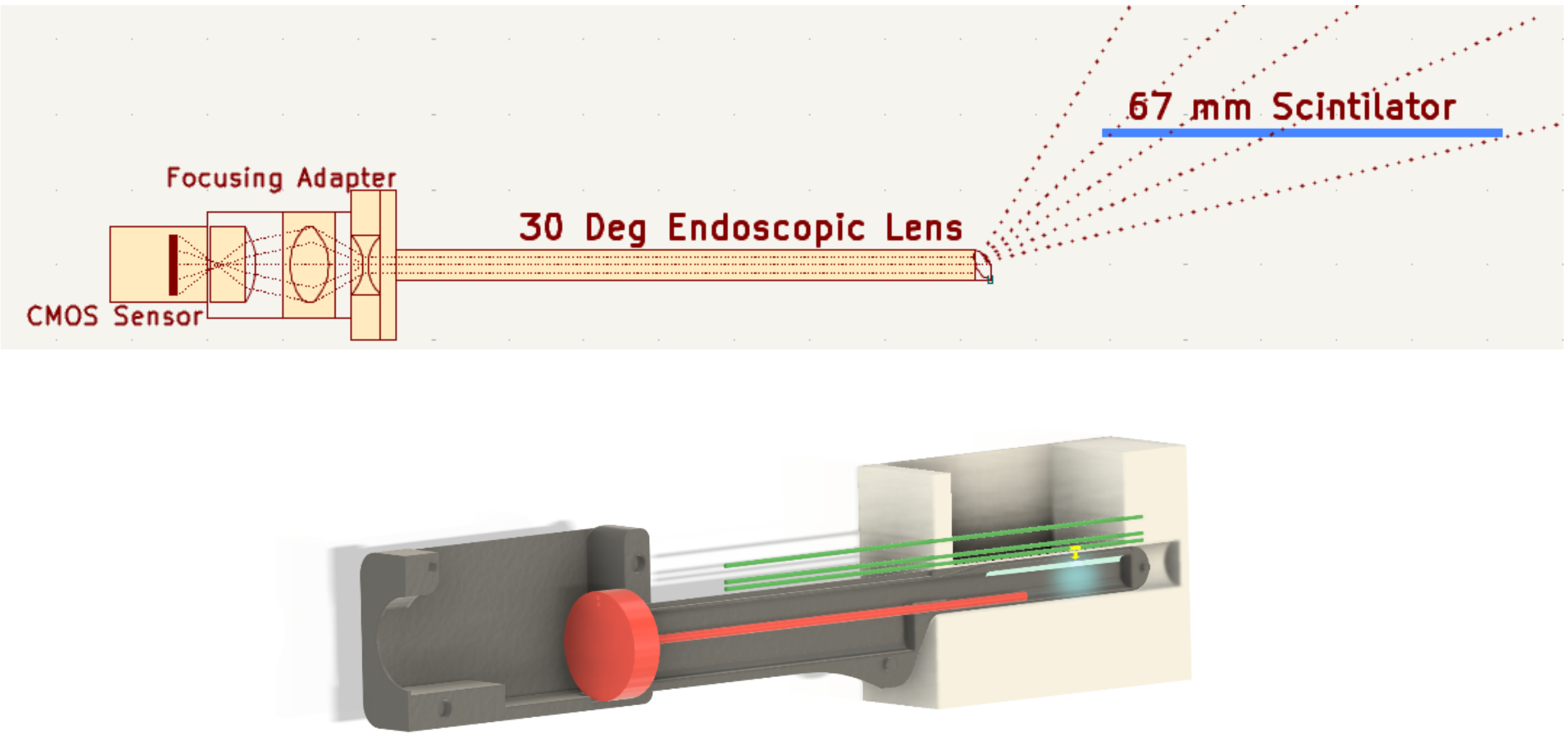

**Figure 1.** Illustration of a cross-sectional view of the dosimeter and test phantom in use. The endoscopic lens is pictured in red, and the 6.7cm scintillator strip is in light blue. In green are 3 interstitial needles positioned to be 7, 10, and 15 mm from the center of the source to the visible surface of the scintillator when in use. The phantom features a cavity for inserting the probe. All components are suspended in water to simulate the tissue environment during use. The small yellow block within the interstitial needle represents the Ir-192 source that illuminates the scintillator below.

### 2.3 HDR delivery test

The system was tested with an HDR delivery using a 10.3 Ci Ir192 source from a Varian Bravos afterloader. The source moved through 3 channels, dwelling 5 seconds at each of five positions spaced 1 cm apart. The system recorded video at 8-bit, 30 frames per second (FPS), with a resolution of 720x540 pixels. The recorded videos were saved as .tiff image stacks for later evaluation.

### 2.4 Image processing

#### 2.4.1 Image geometric and intensity correction

Dosimeter images were denoised using a 3x3 pixel median filter. The denoised images were geometrically corrected using a projective transformation that addresses 8 degrees of freedom: translation (2 degrees), rotation, scale (2 degrees), shear, tilt, and roll. To construct a transformation matrix, the four corners of the imaged scintillator were selected as moving control points to map onto a shell image of the scintillator's physical size (67mm by 5mm) at 96 pixels per inch. Empirical intensity correction factors were derived by summing averaged images at each dwell position. This image was reduced to a one-dimensional intensity profile by computing the median intensity of each column of the image matrix, then smoothing it with a Savitzky–Golay filter using a 3rd-degree polynomial and a 7-pixel frame width.

Profile peaks at the 5 dwell positions in each channel were identified by selecting the maximum value within spatially segmented image regions, yielding a set of variables containing a peak intensity for each dwell position. Since all dwell positions had equal dwell times (t = 5 seconds) and equal source-to-scintillator distances, the target intensity for each dwell was set to the mean of the peak intensities. Then, individual correction factors were computed for each peak as the ratio of the target intensity to the measured peak intensity. With 5 individual correction factors, a 2nd-order polynomial fit obtained an intensity correction function along the length of the scintillator.

**2.4.2 System Calibration**

**2.4.2.1 Calibration of pixel pitch**

Using the same peaks for the intensity correction, the pixel pitch of the image was determined by measuring the distance between adjacent peaks in pixels and relating it to the corresponding 1 cm physical distance. The 4 distances were averaged to yield the average pixel pitch in *mm/pixel.*

$$Pixel\ pitch = \frac{10\ mm}{\overline{N_{pixels}}} \qquad \text{Equation 1}$$

**2.4.2.2 Calibration of Imaged Intensity to Dose**

To find the dose delivered to the scintillator, the system was modelled in Monte Carlo simulations using TOPAS MC.[16,17] This addresses the lack of backscatter from the air cavity beneath the scintillator, which is otherwise ignored by the traditional TG-43[18] method. A model of the GammaMed Plus $^{192}$Ir source was validated by simulating TG-43 conditions and observing agreement within 5% between the TG-43-calculated dose and the simulation results.

This model was used to simulate the dosimeter under two conditions: with an air cavity beneath the scintillator and with that same cavity filled with water.

The model was then used to simulate the expected dose rate for each source-to-scintillator distance, yielding the maximum dose rate and the dose-rate profile within the dosimeter. The maximum simulated dose rate, 21.15 cGy/sec, was then used as the calibration level to relate dose rate to the maximum pixel intensity per frame. This single-point global normalization was validated through two independent checks: cross-depth agreement and consistency in the shape of the dose-rate profile. Cross-depth agreement was assessed by simulating the dose profile delivered to the scintillator at a central dwell position for each of the 3 channel distances and then comparing the simulated and measured profiles. The measured profiles were obtained by extracting the averaged line profile from individual frames captured during a central dwell of scintillator emission, with each frame averaged to reduce noise before comparison.

**2.5 Characterization of Geometric Response Stability**

To determine if source position relative to the scintillator affected the light collected by the camera, the emission at different locations across its surface was measured. The scintillator emission was characterized over an 11-position dwell sequence that spanned the entire scintillator length. In this test, the source positions were placed without spacing between them. The system's response was sampled using the three interstitial needle channels radially offset by 7, 10, and 15 mm from the scintillator.

For each dwell position, a relative emission gradient was established by performing a linear regression of the normalized pixel intensities against the known longitudinal distances. This gradient served as a metric for measuring longitudinal spatial invariance under changing dose rates. The geometric sensitivity variation was then calculated as the coefficient of variation (CV) of these gradients across the stable longitudinal core (Dwells 2–5). This characterization was required to verify that the detector response remains independent of the source's longitudinal position, a prerequisite for employing a single global calibration factor in volumetric dosimetry.

**2.6 Treatment Data Collection and Processing**

**2.6.1 Source Tracking**

One of the key goals of this dosimeter is source tracking. To accomplish this, the physical source position is extracted longitudinally along the scintillator for each frame. The original video, as in 2.4.1, is first geometrically corrected using a projective transformation, resulting in a 105x1407 px image containing only the scintillator. The mean of each column in the video frames is calculated to make a 1-D intensity profile. The profile is smoothed using a cubic spline (smoothing parameter = 0.1). Each frame then has the source position identified by defining the centroid of the full-width-half-max for the intensity profile. This source position matrix then has a 10-frame temporal median filter applied to suppress frame-level outliers. This process results in a matrix of the source position by pixel location at a given time.

#### 2.6.2 Velocity Calculation Framework

The 30 FPS dosimeter position signal was derived from the centroid of the scintillator emission FWHM. Source motion was isolated using a 13 px/frame velocity threshold, and interdwell velocity was calculated by dividing the known 1 cm displacement by the time spent above this threshold within predefined frame regions of interest.

#### 2.6.3 Source Tracking Validation

To validate the source-tracking results, the treatment plan was repeated outside the test phantom using a clear interstitial needle attached to a ruler. The validation experiment used 90 FPS for better temporal resolution, while the dosimeter tests used 30 FPS to allow a longer shutter time (3ms) and a higher signal-to-noise ratio (SNR). Unlike the source tracking in the dosimeter, this tracking was done by first drawing 5 scan lines parallel to the needle's longitudinal axis, spaced ±2 pixels from the center line. The intensity was sampled along each line using bilinear interpolation at 625 points. A 30-sample Gaussian-weighted moving average per line (~3 mm spatial smoothing) was applied. The source position is then found by identifying the location of the maximum absolute intensity gradient, and the final position per frame is the median of the 5 maximum-gradient positions across.

Interdwell segments were identified manually by frame range, and velocity was computed directly from frame-to-frame displacement of the detected edge position. While a frame rate of 30 FPS was used for the dosimeter to maximize the camera’s integration time and SNR, the adequacy of this sampling rate was validated against the high-speed (90FPS) reference. An error analysis was performed, ensuring that the temporal resolution is sufficient to capture the source’s kinetic profile without aliasing or centroid-displacement errors.

To assess the concordance between the 30 FPS dosimeter and the 90 FPS validation testing, a Bland-Altman analysis was performed comparing interdwell velocity measurements. This method quantifies the agreement between two measurement techniques by comparing the difference between paired measurements against the mean of the pair.

The bias was calculated as the mean of the differences, representing the systematic error between the two systems. The 95% limits of agreement (LOA) were defined as:

$$Bias \pm 1.96 \times SD_{diff} \quad \text{Equation 2}$$

Where $SD_{diff}$ is the standard deviation of the differences. This analysis was used to determine if the 30 FPS sampling rate introduced any statistically significant systematic lag or variation compared to the 90 FPS reference.

#### 2.6.4 Dwell time measurement

Dwell Identification: Stationary periods were defined as frames not classified as source motion. A dwell event was recorded when the source remained stationary for a continuous duration exceeding 0.5 seconds (>15 frames at 30 FPS).

Temporal Calculation: The duration of each dwell ($T_{dwell}$)was calculated by dividing the total number of stationary frames ( $N_{frames}$) by the camera sampling rate (f = 30 FPS):

$$T_{dwell} = \frac{N_{frames}}{f} \quad \text{Equation 3}$$

#### 2.6.5 Dose Measurement

Applying the previously established calibration of pixel intensity to dose from the Monte Carlo simulation, the data from this system were converted into a dose-rate profile as a function of source position. A plan of 5 dwell positions spaced 1 cm apart for 5 seconds each was administered to the test phantom using a 10.3 Ci Ir192 source. Each measured dose rate profile, multiplied by the frame duration (1/30 second), yields a dose. The cumulative dose is then calculated by integrating these profile frames over the desired time intervals. This method was utilized to isolate three distinct components:

1. Total Dose: The cumulative dose from the start to the end of the treatment

2. Dwell Dose: The dose accumulated exclusively when the source was stationary at programmed dwell positions.
3. Transit Dose: The dose contribution from the source while in motion between positions.

To validate the system, the measured integral dwell dose was compared against a spatially matched Monte Carlo simulation. To address spatial specificity, a 1-mm point of interest was defined at 325–326 mm along the detector axis, corresponding to a fixed anatomical reference location (analogous to a rectal wall point). The integral dwell dose at this location was extracted from both the dosimeter measurement and the spatially matched Monte Carlo simulation for direct point-wise comparison. Transit dose was isolated at this point to quantify its contribution relative to the dwell dose at a clinically relevant position.

## 3 Results

### 3.1 Intensity Correction Profile and Pixel Pitch Determination

The correction profile[Fig. 2] has a correction factor range magnitude of 1.17. This accounts for the difference in intensity observed in the flat-field tests due to vignetting, without exceeding the expected maximum variation determined by the flat-field analysis.

The same peaks used for the intensity corrections are also used to measure the pixel pitch by relating pixel positions to the planned dwell positions from the afterloader. The system demonstrated a mean pixel pitch of 0.163 mm/pixel (SD: ±0.02 mm). The standard deviation of the position measured was +/-1.3 mm in real-time, and +/-0.8mm retrospectively after a 3-frame moving mean smoothing filter was applied to the position measurements.

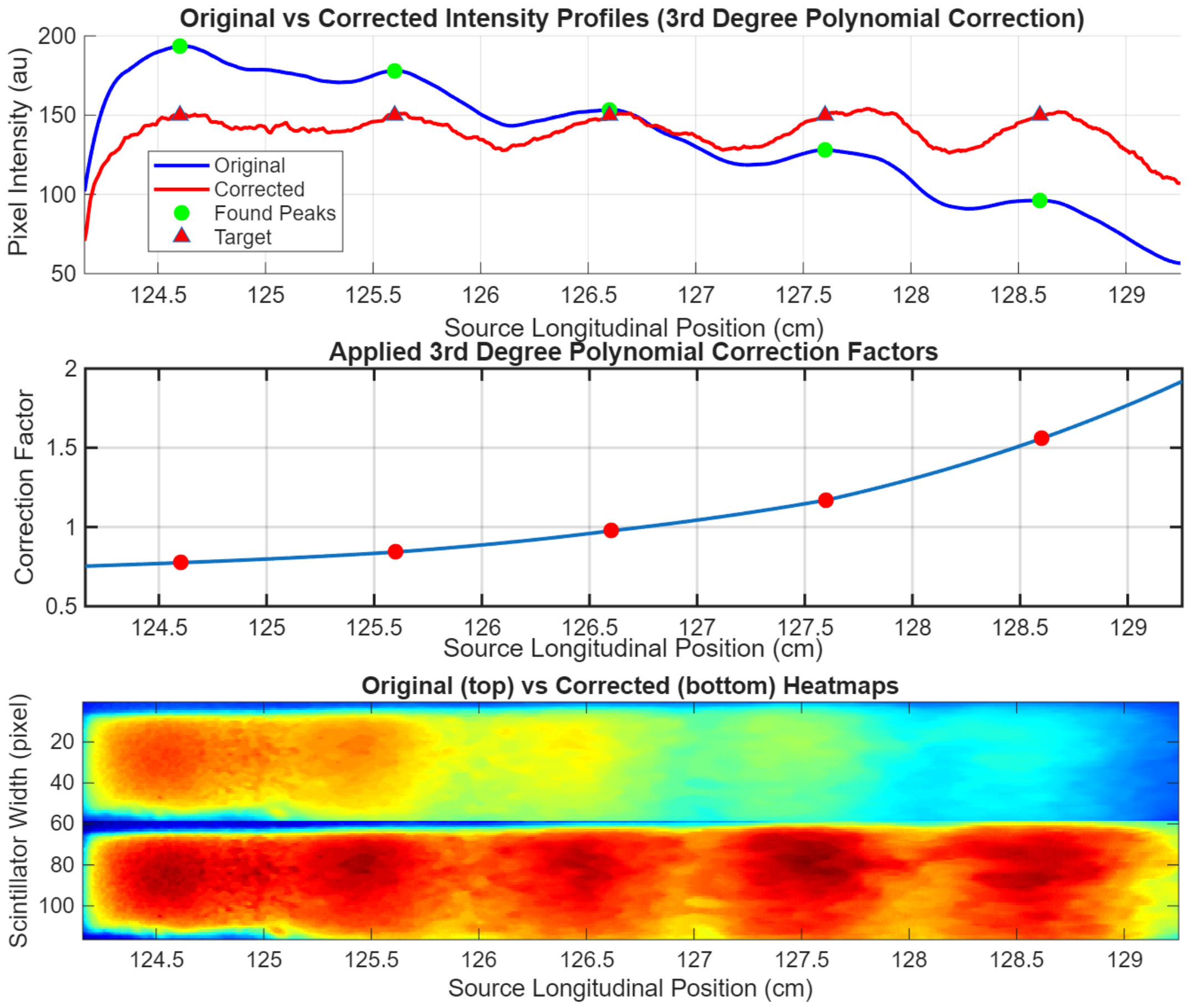

**Figure 2.** Multi-stage calibration process for intensity correction and spatial mapping. Top: longitudinal intensity profiles showing the raw signal (blue, circles) vs. the corrected signal (red, triangles) with the “Found Peaks” annotated. Middle: the applied 3$^{rd}$ degree polynomial correction curve to correct lens vignetting. Bottom: comparative heatmaps of the scintillator before(top) and after(bottom) correction.

### 3.2 Calibration and Monte Carlo

To calibrate and validate experimental measurements, a TOPAS Monte Carlo model was developed to account for the unique geometry of the transrectal probe. Specifically, the model included the impact of the air-filled cavity within the probe, which lacks the backscatter provided by a homogeneous water environment. The simulation revealed a 3.75% decrease in dose rate at the scintillator surface when compared to TG-43 water-equivalent conditions. The peak measured dose rate of 21.15 cGy/sec from the Monte Carlo model was used to derive a normalization factor that converts the maximum pixel intensity to dose rate. This globally applied factor shows pointwise dose-rate agreement at the 10 mm and 15 mm offsets of 7.2% and 0.5%, respectively.

With the measured dose-rate profiles [Fig. 3] calibrated to absolute units (cGy/s) using a single-point normalization factor derived from the peak Monte Carlo dose rate, the first two channels showed excellent spatial agreement with the plan. In contrast, the third channel exhibited a 3 mm longitudinal shift relative to the expected dwell positions. This discrepancy was subsequently confirmed via CT imaging to be a physical displacement of the needle during phantom setup. The system's ability to resolve this 3 mm offset, verified by independent imaging, demonstrates its high sensitivity to common clinical setup errors and its potential as a robust tool for in vivo treatment verification.

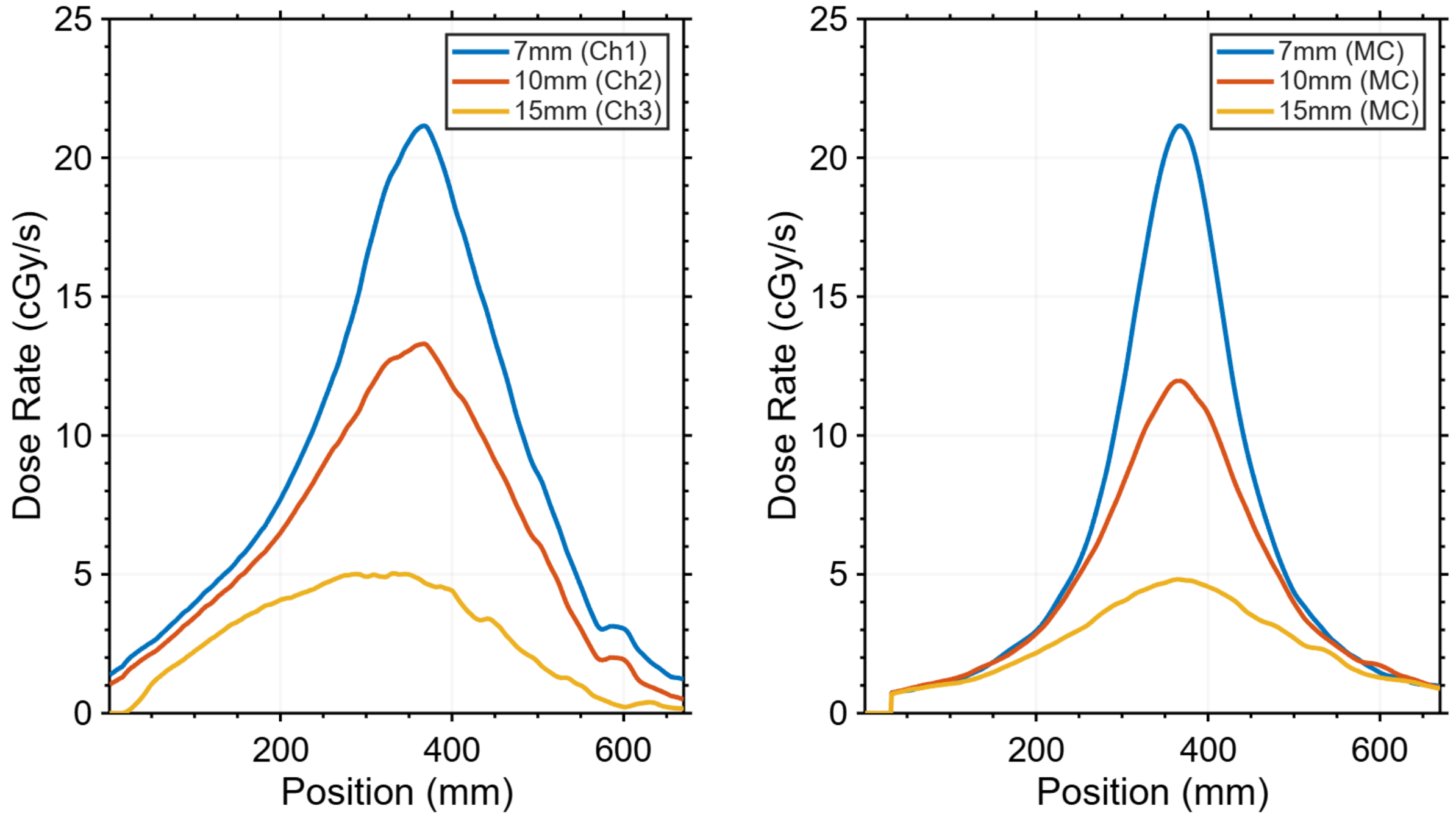

**Figure 3**. Comparison of measured dose-rate profiles (left) and Monte Carlo simulation results (right). The profiles show the 3 separate dose rates measured, highest being the closest to the scintillator (7mm) and lowest being the farthest (15mm), high concordance in peak magnitude and curve shape. Note the 3 mm longitudinal shift in the 15mm experimental channel compared to the simulation, which the system correctly identified as a physical setup error (needle displacement).

**3.3 Geometric Response and Spatial Invariance**

The system demonstrated high linearity across the dose rates corresponding to the three-needle array (7, 10, and 15 mm radial offsets), with all dwell positions within the centrally located dwell positions (Dwells 2–5) achieving a coefficient of regression. $\mathbf{R^2} \geq\ \mathbf{0.995}$. The global mean emission gradient of $\mathbf{0.317}\ \boldsymbol{normalized\ intensity/mm}$, represents the average rate at which measured light intensity changes per unit of longitudinal position along the scintillator, pooled across all dwell positions.

Statistical analysis of the dwell-by-dwell response revealed a geometric sensitivity variation of $\pm\mathbf{2}\%$ across the length of the scintillator [Fig. 4]. While a localized deviation was observed at the initial dwell position (Dwell 1) due to boundary-vignetting effects at the edge of the CMOS field of view and a low signal-to-noise ratio, the response remained highly uniform throughout the central treatment volume. This consistency confirms the spatial invariance of the detector sensitivity across varying dose rates, thereby validating that the calibration factor remains independent of the source's longitudinal position.

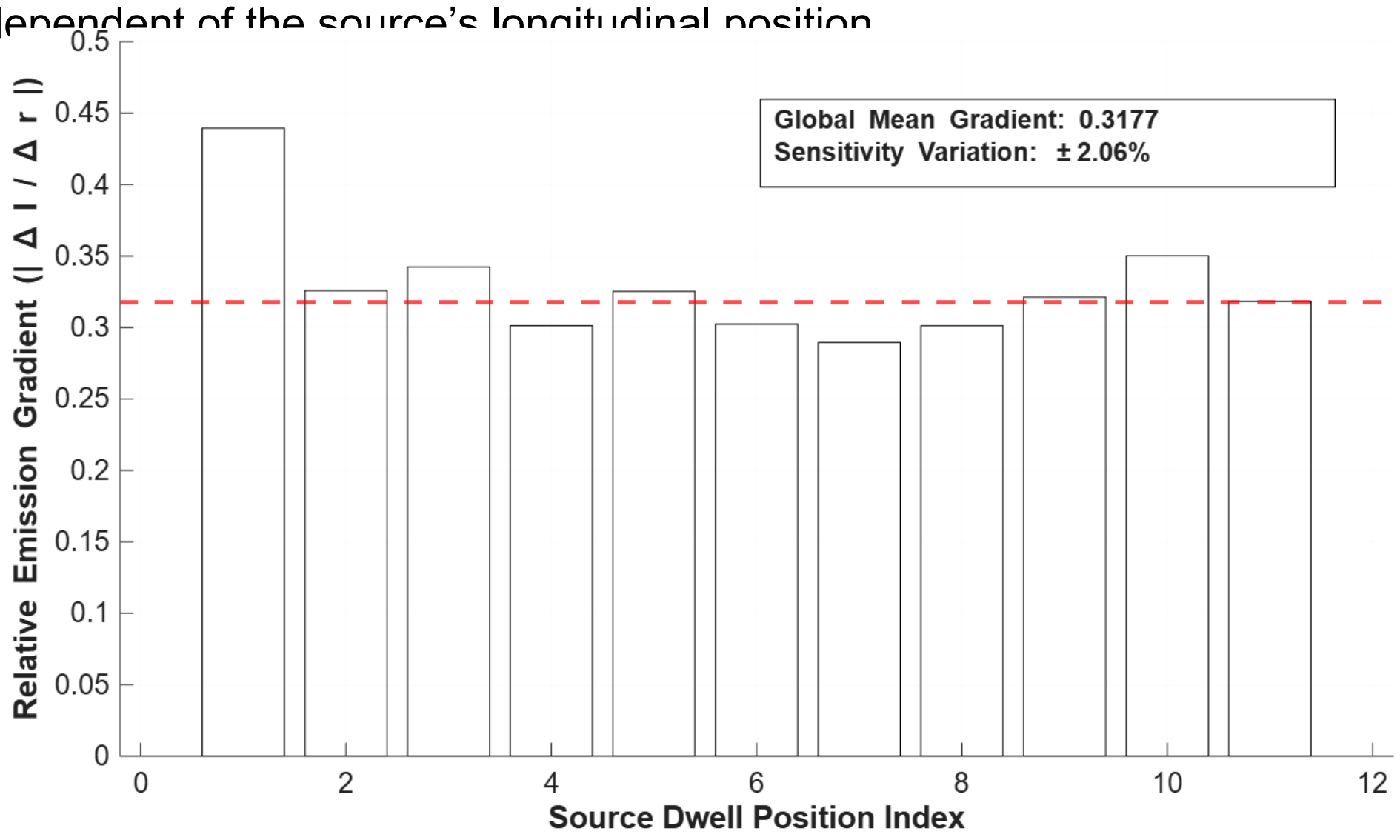

**Figure 4**. Characterization of the scintillator emission gradient across 11 source dwell positions. The gradient represents the linear fit of light intensity captured from three interstitial needles (radial offsets: 7, 10, and 15 mm). High linearity $(\mathbf{R^2} \geq\ \mathbf{0.995})$ was maintained throughout the primary treatment length, with a measured sensitivity variation of $\pm\mathbf{2.06}\%$. This confirms that the camera-based collection system provides a spatially invariant response to varying dose rates across the clinical volume, with the primary deviation observed at the distal boundary dwell position (Dwell 1), attributed to vignetting at the edge of the CMOS field of view, reducing the SNR.

### 3.4 Source Speeds and Dwell Time

The mean interdwell transit speed measured with the 30 FPS dosimeter was 6.06 ± 0.69 cm/s, compared with 6.33 ± 0.43 cm/s obtained using 90 FPS high-speed imaging. Method agreement was assessed using a Bland-Altman analysis, which yielded a mean bias of -0.226 cm/s (95% LOA: -2.23 to 1.78 cm/s). These results indicate high concordance between the two tests. Furthermore, the peak instantaneous source velocity was captured during the initial 'Transit-In' phase; using a 3rd-order polynomial derivative of the 90 FPS position data, a peak velocity of 69.12 cm/s was recorded, well below the manufacturer's nominal specification of 100 cm/s. [Fig. 5]

Average dwell time, measured as the frame count between transits, was 5.02s (SD: 0.0356), slightly higher than the 5.00 s average reported by the clinical afterloader.

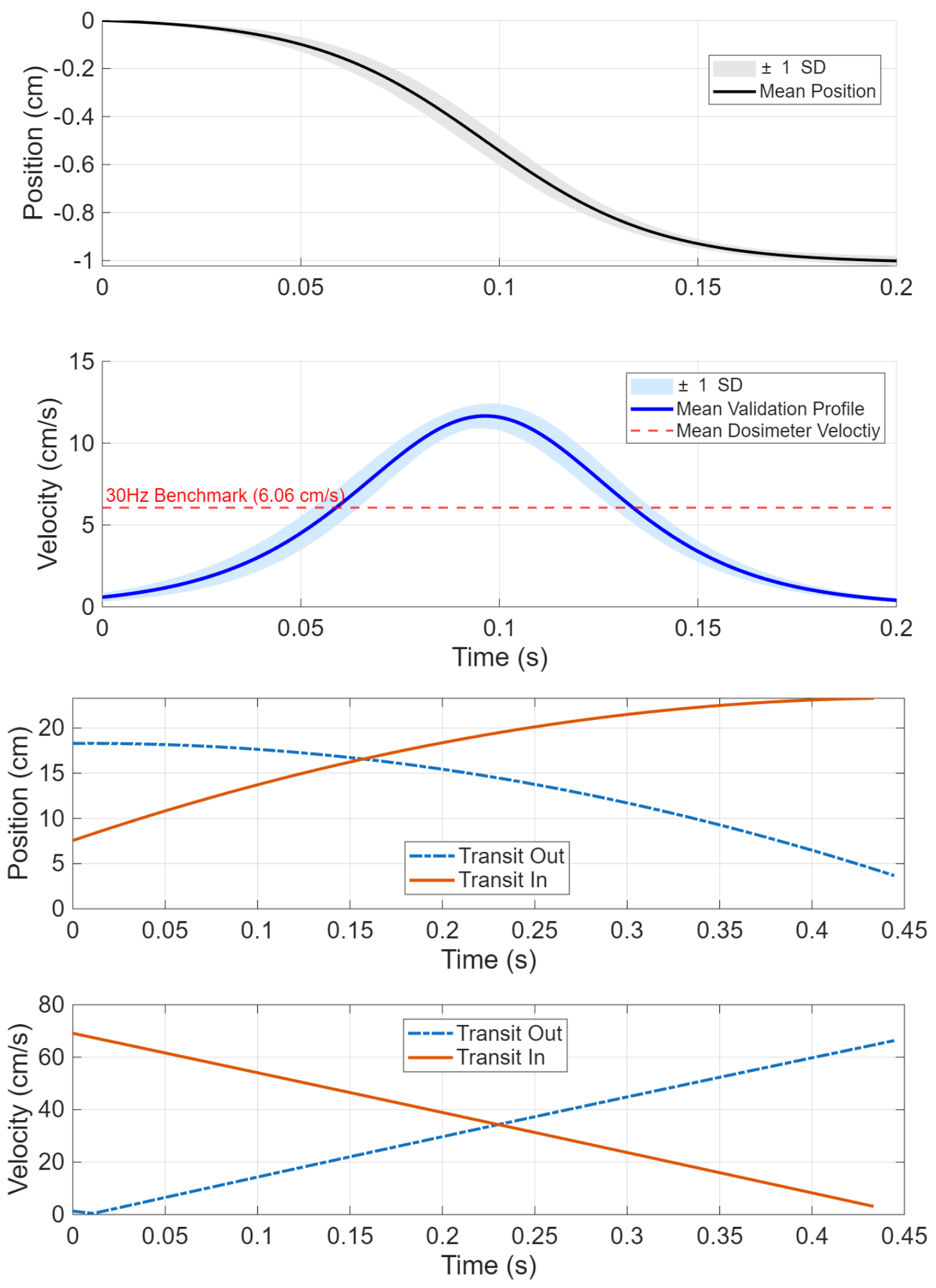

**Figure 5.** Kinematic analysis of HDR source motion for tracking validation. Top panel: Mean longitudinal position (cm) over a 1 cm interdwell jump (0.2 s window). The shaded area

represents ±1 standard deviation across all recorded interdwell transits of this step size, reflecting the reproducibility of the afterloader motion. Second panel: Corresponding velocity profile (cm/s). The solid blue line shows the mean validation profile (90 FPS), while the dashed red line indicates the benchmark velocity (6.06 cm/s) measured by the 30 FPS dosimeter. Third panel: Long-range position tracking for “Transit-In” (source entering needle) and “Transit Out” (source returning to afterloader). Bottom panel: Velocity profiles for long-range transits with a peak of 69 cm/s. The transit-out velocity profile peaks at 69 cm/s within the camera's field of view; the source may continue to accelerate beyond this range.

**3.5 Dose Measurements**

Global validation: The total measured integral dwell dose across all dwell positions was 15.62 Gy, compared to a simulated value of 15.48 Gy, agreeing within 0.9%.

Point-of-interest validation: Simulating a dose measurement to a single point on the anterior rectal wall, a 1mm reference point (325–326 mm), is measured from both the dosimeter and monte carlo data. The integral dwell dose was 319.53 cGy·mm, compared with a Monte Carlo-simulated value of 320.27 cGy·mm, agreeing within 0.23%. The total measured dose at this point, including transit dose, was 340.03 cGy·mm, indicating a transit dose contribution of 20.50 cGy·mm (6.03% of total dose at this location). This is consistent with the global transit contribution of 12.3% observed across the full plan, with the lower value at this specific point reflecting its position relative to the dwell time distribution [Fig. 6].

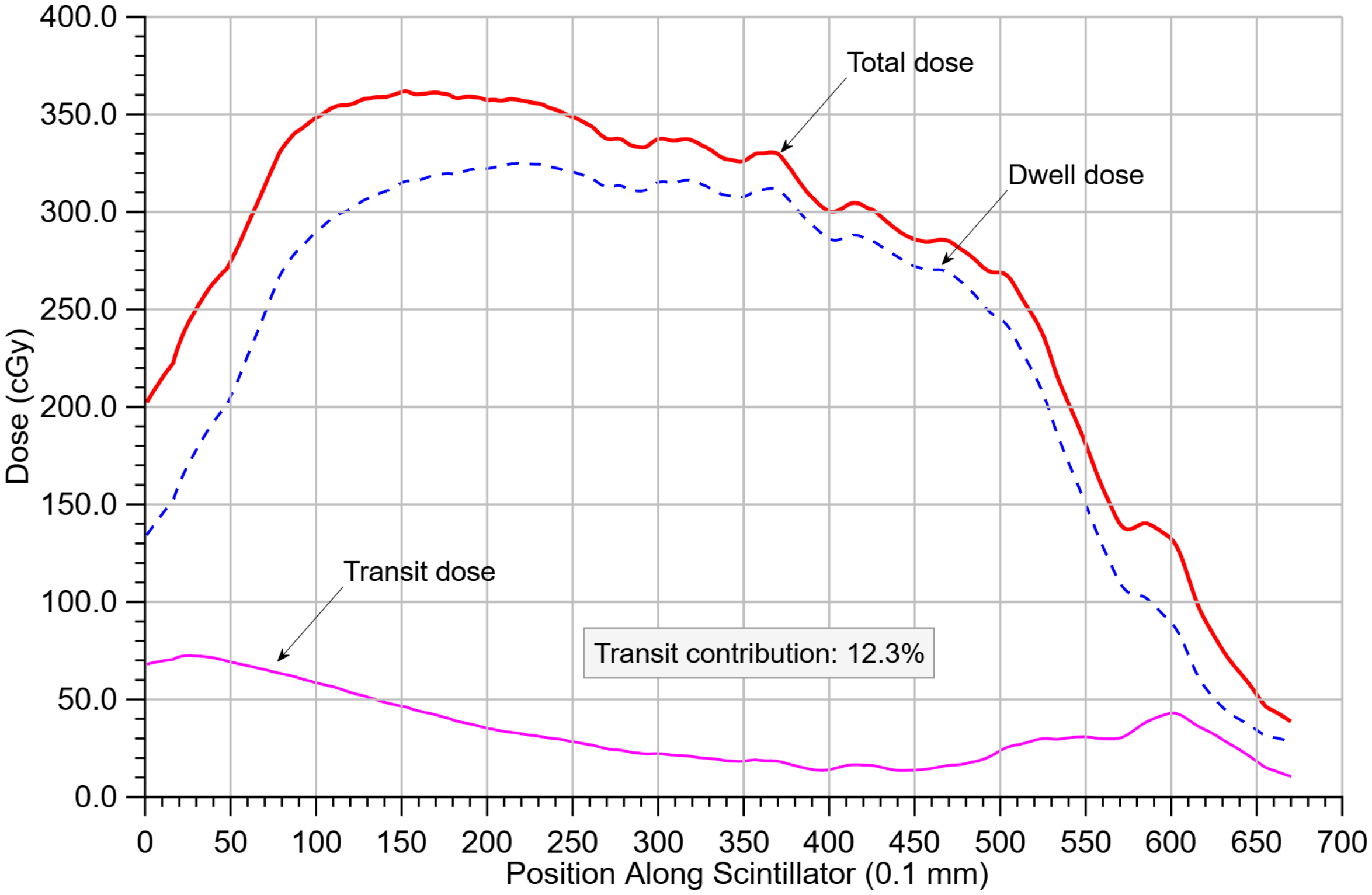

**Figure 6**. Cumulative dose profile along the scintillator for a multi-dwell HDR brachytherapy plan. The plot illustrates the spatial distribution of dose (cGy) as a function of position along the scintillator (units of 0.1 mm). Dwell dose (blue dashed line) represents the dose delivered solely while the source is at its programmed dwell positions. Transit dose (magenta solid line) represents the "extra dose" delivered while the source is moving between dwells. Total dose (solid red line) is the summation of dwell and transit components.

## 4. Discussion

This study demonstrated a scintillator-based optical dosimetry system capable of spatially resolved, real-time HDR brachytherapy verification. The system achieved a spatial sensitivity variation of ±2% across the 67mm field of view, fine spatial resolution of ±1.3 mm in real time and ±0.8 mm retrospectively, and integral dose agreement within 0.9% of the Monte Carlo simulation. Source transit speeds measured at 30 FPS showed a mean bias of -0.226 cm/s relative to the 90 FPS reference imaging, confirming adequate temporal resolution for kinematic characterization. Notably, the system retrospectively resolved a 3 mm needle displacement, subsequently confirmed by CT, demonstrating sensitivity to clinically relevant setup errors.

A critical finding of this work is the substantial contribution of the source transit to the total delivered dose. In the evaluated 5-dwell proof-of-concept plan, the transit dose accounted for 12.3% of the total integral dose. This contribution is highly dependent on source strength, the number of dwell positions, and interdwell distances. Failure to account for these interdwell doses in clinical planning software presents a risk of systematic overdose from the prescribed doses. This error decreases in relative magnitude as the source decays because of less relative time spent in transit compared to that spent at dwell positions.

### 4.1 Technical Considerations and System Refinements

The prototype exhibited reduced accuracy and a lower SNR at the farthest end of the measurement surface. Future iterations may mitigate this by applying reflective coatings or specialized paints to the scintillator edges to enhance total internal reflection and maximize light collection toward the detection lens.[12,19]

The prototype exhibited an increased emission gradient at the initial boundary of the measurement volume (Dwell 1, as seen in Fig. 3). This localized distortion is primarily attributed to light leakage at the proximal scintillator end facet, where uncoated boundaries allow optical emission to escape laterally rather than propagate entirely toward the collection lens. This escaping light steepens the local spatial rate of change of intensity, distorting the gradient. Future iterations can mitigate this edge effect by applying reflective coatings or specialized optical paints to the scintillator end facets, enhancing total internal reflection and maximizing uniform light collection toward the detection lens.

The observed 7.6% pointwise discrepancy at the 10 mm offset is a primary limitation of the current prototype. While the scintillator material itself possesses high intrinsic resolution, the optical housing design introduces significant spatial broadening. Light generated in the scintillator is not perfectly baffled, causing internal reflections (optical crosstalk) within the probe housing before it reaches the lens. This creates a spatially symmetric "halo" effect that widens the detected peak relative to the Monte Carlo scoring geometry. Because this broadening is symmetric, the total light collected remains proportional to the energy deposited, which explains why the integral-dose agreement remains within 0.9% despite the pointwise "blurring." Future iterations must utilize internal optical baffling or blackened channels to suppress these reflections and improve spatial fidelity.

The scope of this study was limited to a proof-of-concept using an idealized phantom design that positioned the needles in the “D” column of a clinical interstitial brachytherapy grid. In transrectal ultrasound imaging, the prostate gland is deformed under the anterior pressure applied by the transrectal probe, and the probe itself is purposefully displaced from the anterior rectal wall by the probe cover to increase the field of view. The central “D” column is typically avoided in treatments, with the needles placed to the left or right of the “D” column to protect the urethra. The response of the scintillator to sources displaced laterally from the central axis (D grid position) remains to be investigated, as does the response from a source lateral and posterior to the scintillator due to prostate deformation. These clinically relevant displacements of the source from the central axis of the scintillator detector may result in inaccurate response of the detector relative to the anterior rectal wall, the magnitude of which is open for further study. These effects could potentially be mitigated via a true 2D curvilinear detector; however, other complex issues such as internal reflections will likely be introduced and need to be accounted for.

### 4.2 Future Directions

A logical progression for this technology is its integration into an ultrasound-equipped probe. The ability to map spatially resolved dose measurements directly onto real-time anatomical imaging would facilitate the generation of in vivo dose profiles, enabling direct comparison of delivered and planned doses at the probe displacement from the rectal wall. However, integration of an ultrasound array and its associated electrical contacts would introduce additional attenuation of the scintillator signal, necessitating recalibration of the dose-to-intensity conversion factor for the modified probe geometry.

The current air-filled cavity beneath the scintillator presents a tradeoff between optical and mechanical performance. While the air interface is a simple model for this prototype, it provides no mechanical support to the optical components. A clear epoxy filler could improve the structural stability of the lens and provide acoustic coupling for an integrated ultrasound transducer. Still, it would require characterization for radiation-induced opacification over the device's clinical lifespan, as progressive darkening would introduce a time-varying attenuation factor that would require routine recalibration.

## 5 Conclusions

This study demonstrates a novel, spatially resolved optical dosimetry system for real-time verification of HDR prostate brachytherapy. By integrating high-speed imaging with a scintillator-embedded transrectal probe, the system achieves high spatial resolution, ±2% sensitivity uniformity, and integral dose agreement within 1% of Monte Carlo simulation. The system successfully resolved a 3-mm needle displacement, which was subsequently confirmed by CT, demonstrating clinically meaningful sensitivity to common setup errors.

The transit dose accounted for 12.3% of the total delivered dose in the evaluated plan. This poses the risk of a systematic overdose contribution if not accounted for in clinical planning software, and of greatest concern for high-activity sources early in their clinical lifespan.
The Monte Carlo analysis confirmed that tissue heterogeneity has minimal dosimetric impact for Ir-192, with even a worst-case 1 cm air cavity producing only a 3.75% dose deviation. This reinforces that the primary sources of clinical uncertainty in HDR brachytherapy are source kinematics and physical setup errors, both of which this system is uniquely positioned to monitor.